\documentclass[epj,final]{svjour}
\usepackage{graphics}

\usepackage{mathtext}
\usepackage{graphicx}
\usepackage{amsmath}
\usepackage{amsfonts}
\usepackage{amssymb}

\begin{document}

\title{Initial-state nuclear effects in proton-nucleus collisions}
\author{Ya.~A.~Berdnikov\inst{1,2} \and 
        V.~T.~Kim\inst{1,2} \and
	V.~F.~Kosmach\inst{1} \and
	M.~M.~Ryzhinskiy\inst{1,2}\thanks{\emph{E-mail:} mryzhinskiy@phmf.spbstu.ru}
	\and
	V.~M.~Samsonov\inst{1,2} \and
	M.~E.~Zavatsky\inst{1,2}
}                     
%
%
\institute{St. Petersburg State Polytechnic University, St. Petersburg, Russia \and 
           Petersburg Nuclear Physics Institute, Gatchina, Russia}
\date{Received: date / Revised version: date}
%
\abstract{
Two important initial-state nuclear effects in hadron-nucleus collisions are considered.
The ratios of inclusive differential cross sections for Drell-Yan dimuon production 
are calculated. The calculated results are compared to the E866 data. It is shown that 
consideration of multiple soft rescatterings of incident quarks in nuclei and initial-state
quark energy loss effects allow to get a good agreement between the calculated results 
and the experimental data.
} 
\titlerunning{Initial-state nuclear effects in pA collisions}
\maketitle
\graphicspath{{./eps/}{./}}
\newcommand{\e}{\mathrm{e}}

\section{Introduction}
\label{intro}
Nowadays there is a great interest among the widest circles of physicists in non-trivial
effects of relativistic nuclear physics such as anomalous nuclear dependence in processes
with large transverse momentum $p_{\text{T}}$. This problem has become especially
important in connection with the recent data from RHIC 
\cite{Arsene:2003yk,Adler:2003au,Back:2003qr,Adams:2003kv}.

It was first observed in 1975 \cite{Cronin:1974zm} that high-$p_{\text{T}}$ hadrons 
in proton-nucleus (pA) collisions are produced copiously in the range of
$p_{\text{T}} \gtrsim 2$~GeV/{\it c}. This effect (``Cronin effect'') was observed in fixed target pA collisions 
at energies 200, 300 and 400~GeV.
Cronin effect demonstrates that 
a hadron-nucleus (hA) collision can not be presented as a simple superposition of 
hadron-nucleon (hN) collisions. Analogous behavior was observed in collisons of heavy 
nuclei (Pb+Pb, Pb+Au) at $\sqrt{s_{\text{NN}}}=17$~GeV (SPS, CERN) \cite{Aggarwal:1998vh}.
But the recent experimental data from RHIC showed strong suppression of produced hadrons
in central Au+Au collisions at $\sqrt{s_{\text{NN}}}=130, 200$~GeV 
\cite{Arsene:2003yk,Adler:2003au,Back:2003qr,Adams:2003kv}.

The anomalous A-dependence can be affected by initial- and final-state effects, {\it i.e.} the effects 
before and after hard scattering respectively. The investigation of the final
state could give the information on the properties of produced medium. But this information can be extracted
from data only when the initial state can be reliably predicted. A unique tool for studying the initial state
is the Drell-Yan (DY) lepton-pair production \cite{Drell:1970wh}, which provides the possibility of probing 
the propagation of partons through nuclear matter in its ground state, with produced lepton pair carrying away
the desired information about the initial state, without being affected by the produced medium.

In this paper two important initial-state effects which take place in hA collisions are studied: 
multiple soft rescatterings of quarks of the incident hadron in nuclei and energy loss of fast quarks in nuclear matter.
In order to simulate the mentioned effects in DY lepton-pair production in proton-nucleus and nucleus-nucleus (AA)
collisons we developed a new Monte Carlo (MC) event generator HARDPING (Hard Probe Interaction Generator). It is based
on HIJING generator \cite{Gyulassy:1994ew}, an extension of PYTHIA \cite{Sjostrand:1993yb} for hadron collisions
on jet production in nuclear collisions.

Multiple soft rescatterings and energy loss of quarks were
taken into account according to refs. \cite{Efremov:1985cu,Johnson:2001xf}. 
Obtained results were compaired with E866 Collaboration (FNAL, USA) data at 800~GeV
\cite{Vasilev:1999fa}.

\section{Studying the initial state}
\label{sec:1}

\subsection{Nuclear shadowing versus energy loss}
\label{sect:shadowing}
Measurements of nuclear structure functions in deeply inelastic lepton-nucleus scattering 
(DIS) \cite{Aubert:1983xm,Arneodo:1996rv,Adams:1995is} indicate cle\-ar\-ly that parton 
distributions of bound nucleons are different from those of free nucleons.

It is very important to disentangle between the effects of shadowing and energy loss, since they 
are similar in many respects. In order to describe the modification of parton distributions in 
nucleus, a variety of approaches to this question exist in literature 
\cite{Amaudruz:1995tq,Geesaman:1995yd}.

The first DY data suitable for such an analysis were obtained in Fermilab E772/E866 experiments.
An analysis of the E772 data \cite{Alde:1990im} was made in ref. 
\cite{Gavin:1991qk}, ignoring shadowing. A better analysis was performed
by the E866 collaboration using the E866 data \cite{Vasilev:1999fa}. The E866 experiment extends kinematic 
coverage of the previous E772 experiment, which significantly increases its sensitivity to parton 
energy loss and shadowing. Vasiliev {\it et al.} attempted to improve the analysis of ref. \cite{Alde:1990im} 
by including shadowing. However the procedure employed by Vasiliev {\it et al.} used the EKS shadowing
parametrization \cite{Eskola:1998df}, which included the E772 DY data, which are subject to corrections
for energy loss. Thus the EKS ``shadowing'' already includes corrections for energy loss. And this is why
the analysis of the E866 data performed by the E866 collaboration resulted in zero energy loss.

In 2001, Hirai, Komano and Miyama (HKM) \cite{Hirai:2001np} proposed nuclear
parton distributions, which were obtained by quadratic and cubic type analysis, and determined
by a $\chi^2$ global analysis  of existing experimental data on nuclear structure functions without
including the proton-nucleus DY process.

In our present analysis we use the HKM nuclear shadowing parametrization. Since the HKM fit did not
include Drell-Yan data, we expect to find energy loss and shadowing corrections which are 
unambiguously separated.

\subsection{Multiple soft rescatterings of quarks}
\label{sect:multrescat}
First of all, the mechanism of multiple interactions significantly changes with energy.
At low energies a high transverse momentum parton is produced off different nucleons 
incoherently, while at high energies it becomes a coherent process. This is controlled by 
the coherence length \cite{Kopeliovich:2002yh}
\begin{equation}
l_c = \frac{\sqrt{s}}{m_\text{N} k_\text{T}},
\end{equation}
where $k_\text{T}$ is the transverse momentum of the parton produced at mid rapidity
and then hadronizing into the detected hadron with transverse momentum $p_\text{T}$.

For a coherence length which is shorter than the typical internucleon separation, the
projectile interacts incoherently with individual nucleons. The energy range of the E866
experiment corresponds to the regime of short coherence lengths. Hence  effects of coherence
are not so important here. Thus we are not going to consider coherence effects in our 
present analysis.

In hA interactions a quark of the incoming hadron can undergo soft collisions (with small
momentum transfers: $|t|<1~\text{GeV}^2$) as well as hard ones (with large momentum
transfers: $|t|>1~\text{GeV}^2$) inside the nucleus. It was first shown by Levin and Ryskin
\cite{Levin:1981mv} that the observed Cronin effect can not be explained by only
hard collisions taken into account. One also have to consider soft rescatterings of additive quarks 
of the incident hadron before the hard process \cite{Voloshin:1982ry}. 

Such a picture was suggested in ref. \cite{Efremov:1985cu} in the framework of the additive quark model
\cite{Anisovich:1985xd}. In this approach the dynamics of hA interactions could be visualized in 
the following manner. Each constituent quark of the incident hadron (valon)
is scattered independently of the other quarks (as in the additive quark model) 
several times softly, {\it i.e.} with small momentum transfer, by the
nucleons of the target nucleus. Then a (anti-) quark-parton, which belongs to this quark, undergoes
a hard collision with an antiquark (quark) of the nucleus producing an observed DY pair. 
All these soft rescatterings affect 
the $p_\text{T}$ distribution of the quarks of the incident hadron, and therefore, the $p_\text{T}$ spectrum
of observed DY pairs.

According to ref. \cite{Efremov:1985cu} the probability for a quark to undergo $n$ soft rescatterings 
in nuclear matter is:
\begin{eqnarray}
  P_n &=& \frac{1}{(n-1)!} \int\limits_{-\infty}^\infty \text{d}z \int \text{d}^2b\, 
  \bigl[ \sigma T_-(b,z)\bigr]^{n-1} \nonumber \\
  \label{eq:pn}
  &&{}{} \times \rho(b,z)\,\e^{-\sigma T_-(b,z)}\ .
\end{eqnarray}

Here $b$ is the impact parameter of hA collision, $\sigma$ is the cross section of the inelastic soft
quark-nucleon interaction, $\rho(r)$ is the nuclear density normalized to unity
\begin{equation}
\nonumber
\int \rho(r)\,\text{d}^3r=1\ ,
\end{equation}
$T_-(b,z)$ is the profile function:
\begin{equation}
\nonumber
T_-(b,z)=\int\limits_{-\infty}^z \rho(b,z^\prime)\text{d}z^\prime\ .
\end{equation}

Transverse momentum $k_\text{T}$ distribution for the quark suffered $n$ soft rescatterings can be written in the following
form \cite{Lykasov:1984uf}:
\begin{equation}
\label{eq:mult_1}
G^n_\text{q}(k_\text{T})=\int \prod\limits_{i=1}^n \text{d}^2p_{\text{T}_i}\,
f_\text{q}(p_{\text{T}_i})\,\delta^2(k_\text{T}-\sum\limits_{i=1}^n p_{\text{T}_i}) \ ,
\end{equation}
where $p_{\text{T}_i}$ is the transverse momentum of the quark in $i$-th rescattering process. Here 
$f_\text{q}(p_\text{T})$ is the probability for 
the quark q to have the transverse momentum $p_\text{T}$ after a single quark-nucleon interaction:
\begin{equation}
f_\text{q}(p_\text{T})=\frac{1}{\sigma}\,\frac{\text{d}\sigma}{\text{d}^2p_\text{T}}\ .
\end{equation}
Finally, eq. (\ref{eq:mult_1}) can be presented in the following form \cite{Alaverdian:1976qh}:
\begin{eqnarray}
  G_\text{q}^m(k_\text{T}) & = & \frac{B^2}{2\pi\Gamma\bigl[(3m+1)/2+1\bigr]}\biggl(\frac{Bk_\text{T}}{2}\biggr)^{(3m+1)/2}\,\nonumber\\
  \label{eq:kim_gq}
  && {} \times K_{(3m+1)/2}(Bk_\text{T})\ ,
\end{eqnarray}
where $m=n-1$, $K^m(y)$ is the McDonald function of $m$ order, $\Gamma(\alpha)$ is
the gamma function, $B=2/\langle k_\text{V} \rangle$, where $\langle k_\text{V}\rangle$ is the mean
momentum of the quark in the projectile proton.

\begin{figure}[htb]
  \centering
  \includegraphics[width=.9\hsize]{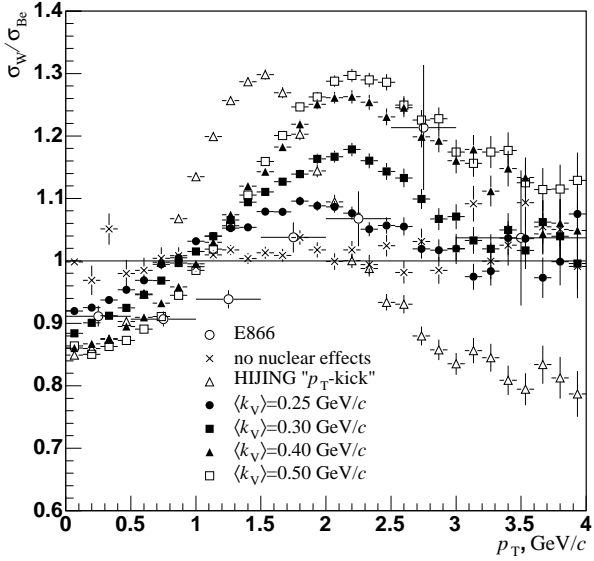}
  \caption{Ratio of the cross sections for Drell-Yan events versus $p_\text{T}$ with quarks multiple
    rescatterings taken into account. Ratios calculated for $\sigma=10$~mb and for different
    $\langle k_\text{V}\rangle$ values.}
  \label{fig:nconv_s10}
\end{figure}

The modification of primordial $p_\text{T}$ due to soft rescatterings of the quark 
undergoing the hard process  was implemented in HIJING according to the mentioned
distributions (eqs.~(\ref{eq:pn}) and (\ref{eq:kim_gq})). Fig.~\ref{fig:nconv_s10} represents
HARDPING simulation results for the ratios of the DY pairs production cross 
sections in pA collisions versus $p_\text{T}$ of produced pairs. The ratios were calculated 
for different $\langle k_\text{V}\rangle$ values and for $\sigma=10$~mb. Also the simulation results, obtained
with the standard HIJING multiple scatterings algorithm (HIJING ``$p_\text{T}$-kick'') \cite{Gyulassy:1994ew}
as well as E866 data are presented.

The ratio $\sigma_\text{W}/\sigma_\text{Be}$ is a fraction of inclusive differential cross sections, normalized to the 
corresponding atomic number:
\begin{equation}
  \sigma_\text{W}/\sigma_\text{Be}=\biggl(\frac{1}{A_\text{W}}\,\frac{\text{d}\sigma^\text{pW}}{\text{d}p_\text{T}}\biggr) \bigg/ 
  \biggl(\frac{1}{A_\text{Be}}\,\frac{\text{d}\sigma^\text{pBe}}{\text{d}p_\text{T}}\biggr)\ ,
  \label{eq:ratio_sigma}
\end{equation}
where $A_\text{W}$ and $A_\text{Be}$ are atomic numbers of corresponding nuclei, $\text{d} \sigma^\text{pW} / \text{d}p_\text{T}$
and $\text{d} \sigma^\text{pBe} / \text{d}p_\text{T}$ are inclusive differential cross sections for DY pairs production
in corresponding reactions.

\begin{figure}[htb]
  \centering
  \includegraphics[width=.9\hsize]{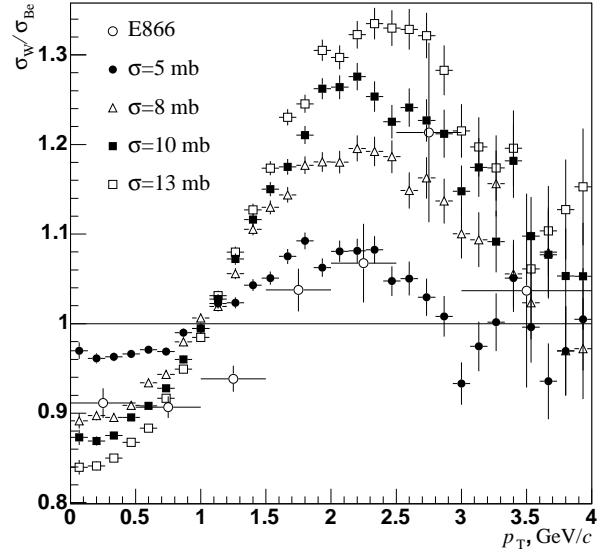}
  \caption{The same as in fig.~\ref{fig:nconv_s10} but for different $\sigma$ and $\langle k_\text{V}\rangle=0.4$~GeV/{\it c}.}
  \label{fig:nconv_k04}
\end{figure}

Fig. \ref{fig:nconv_k04} represents the same ratio for different values of quark-nucleon cross 
section and for $\langle k_\text{V}\rangle=0.4$~GeV/{\it c}.

\begin{figure}[htb]
\centering
\includegraphics[width=.9\hsize]{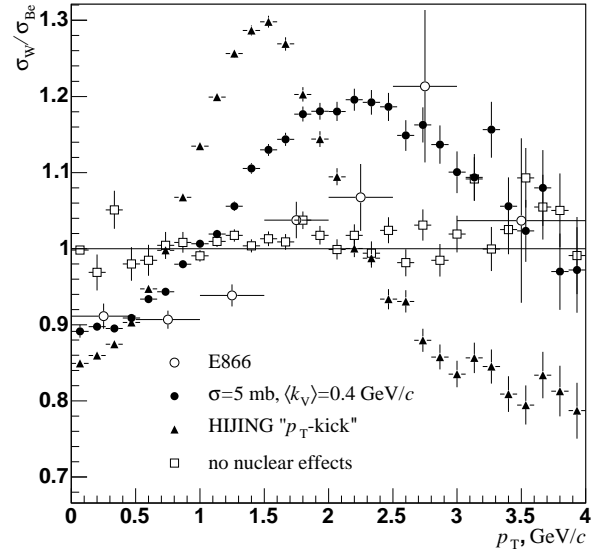}
\caption{The same as in fig.~\ref{fig:nconv_s10} but for $\sigma=5$~mb and 
$\langle k_\text{V}\rangle=0.4$~GeV/{\it c}.}
\label{fig:nconv_optimal}
\end{figure}

As seen from figs.~\ref{fig:nconv_s10} and \ref{fig:nconv_k04} considering the multiple soft quark 
rescatterings effect allows to improve agreement with the experimental data comparing to the results 
obtained without taking into account this effect as well as to the HIJING ``$p_\text{T}$-kick'' results. Also one can
see the results strongly depend on the values of $\sigma$ and $\langle k_\text{V}\rangle$. Unfortunately, in spite of the obvious
improvement of agreement between experimental and simulated data comparing to the original HIJING,
there is a large quantitative inconsistency. In fig.~\ref{fig:nconv_optimal} one can
see the best result obtained within this model, where the value of $p_\text{T}\approx 2.2$~GeV/{\it c}, 
corresponding to the maximum of the ratio, is about $1.2$ times less than the experimental one.
Moreover, the value of $\sigma=5$~mb corresponding to this figure is about 2 times less than the
one predicted by the additive quark model ($\sigma\approx \frac{1}{3}\,\sigma_\text{NN}\approx 10$~mb,
where $\sigma_\text{NN}\approx 30$~mb). So we need more accurate model.

The above treatment of multiple scatterings of quarks was oversimplified because there
was no distinction between constituent quarks (valons) which undergo soft scatterings and
point-like partons which undergo hard process. In fact, to obtain distribution function of
the parton, which undergoes the hard process, of the incident hadron one should calculate
the convolution of the distribution function of a valon after $n$ multiple soft
collisions $G_\text{q}^n(k_\text{T})$ and the distribution function of the parton inside the 
valon $F_\text{V}(k_\text{T})$:
\begin{eqnarray}
  F_\text{N}(k_\text{T}) &=& G^n_\text{q}\otimes F_\text{V} \nonumber\\
  & = & \int \text{d}^2p_{\text{T}_1} \text{d}^2p_{\text{T}_2}\, 
  G^n_\text{q}(p_{\text{T}_1})F_\text{V}(p_{\text{T}_2})\, \nonumber\\
  \label{eq:kim_fN}
  & & {}\times \delta^2(k_\text{T}-p_{\text{T}_1}-p_{\text{T}_2})\ ,
\end{eqnarray}

The distribution function of the parton inside the valon was taken in the
following form:
\begin{equation}
\label{eq:f_q}
F_\text{V}(k_\text{T})=\frac{B^2}{2\pi}\,\e^{-Bk_\text{T}}\ ,
\end{equation}
where $B=2/\langle k_\text{q}^\text{hard} \rangle$, and $\langle k_\text{q}^\text{hard}\rangle$ is the mean
momentum of the parton in the valon.

\begin{figure}[htb]
\centering
\includegraphics[width=.9\hsize]{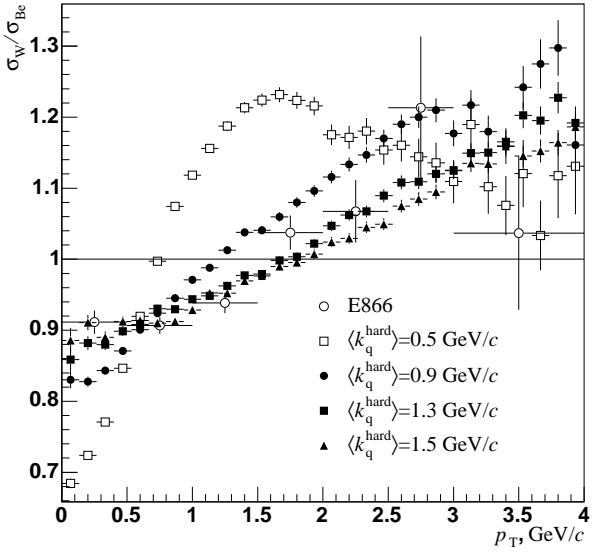}
\caption{Ratio of the cross sections for Drell-Yan events versus $p_\text{T}$ calculated for 
different $\langle k_\text{q}^\text{hard}\rangle$ and $\langle k_\text{V}\rangle=0.4$~GeV/{\it c} (see the text).}
\label{fig:yconv_ks04}
\end{figure}

\begin{figure}[htb]
\centering
\includegraphics[width=.9\hsize]{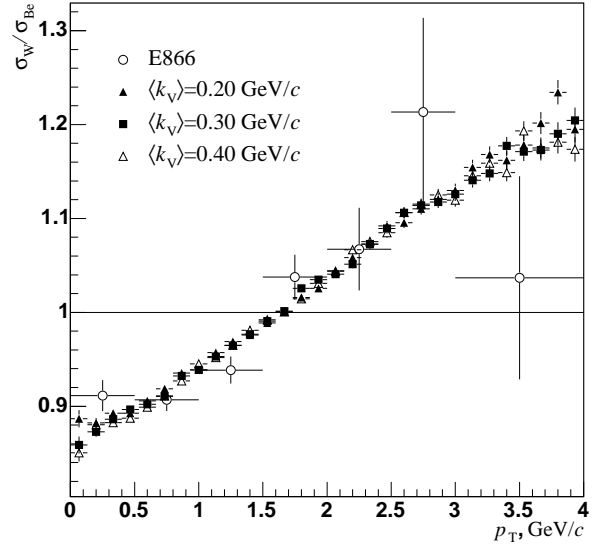}
\caption{The same as in fig.~\ref{fig:yconv_ks04} but for different $\langle k_\text{V}\rangle$ and 
$\langle k_\text{q}^\text{hard}\rangle=1.3$~GeV/{\it c} (see the text).}
\label{fig:yconv_kh13}
\end{figure}

The HARDPING calculation results with the above treatment of multiple interactions are
presented in figs.~\ref{fig:yconv_ks04} and \ref{fig:yconv_kh13} for
different values of $\langle k_\text{q}^\text{hard} \rangle$ and $\langle k_\text{V} \rangle$. All
calculations were performed for $\sigma=10$~mb, which corresponds to the additive quark model.
One can see that the ratio $\sigma_\text{W}/\sigma_\text{Be}$ is rather sensible to 
the variations of $\langle k_\text{q}^\text{hard} \rangle$  and changes weakly with the variations
of $\langle k_\text{V} \rangle$. The new model gives us much better agreement with the 
experiment than the previous one. One should estimate the optimal model parameters as
\begin{eqnarray*}
\sigma & = & (8 - 10)~\text{mb}, \\ 
\langle k_\text{q}^\text{hard} \rangle&=&(0.8 - 1.5)~\text{GeV/{\it c}}, \\
\langle k_\text{V} \rangle&=&(0.2 - 0.4)~\text{GeV/{\it c}}.
\end{eqnarray*}

\subsection{Energy loss of fast quarks in nuclear matter}
\label{sect:eloss}
There is another important process which affects proceeding of hadron-nucleus
collisions -- the enery loss of fast quarks while travelling through
nuclear matter. This process also takes a great interest last years, there are
a lot of models which are trying to describe such an effect, see {\it e.g.} ref.
\cite{Garvey:2002sn}, but unfortunately there is no general agreement on the value
of the partonic energy loss rate $\text{d}E/\text{d}z$ neither in cold nor in hot nuclear matter.

Much of the problem originates from the impossibility of direct measurements of this
energy loss. And the lack of a common agreement of the processes and mechanisms to
be included in specifying the energy loss leads to large differences between the
results of each separate experiment. The energy loss in the initial state are 
usually measured from the data on A-dependence of the Drell-Yan pair production
in proton-nucleus collisions, see {\it e.g.} ref. \cite{Vasilev:1999fa}.  In this paper
we adopted the model for initial-state quark energy loss described in ref. \cite{Johnson:2001xf}.

It is usually assumed that quark propagates from the surface of the nucleus
to the point where the DY pair is produced, which would mean that the mean quark
path in the nucleus would be $\langle L\rangle\approx 3R_\text{A}/4$. But, as shown
in ref. \cite{Kopeliovich:1984bf}, this value should be shortened by at least the mean free path
of a proton in a nucleus, $\approx 2$~fm. This would substantially reduce 
$\langle L\rangle$ by a factor of two or more, so that the mean path between 
the point of DY pair production and the first inelastic interaction is actually shorter
than the maximum possible distance to the edge of the nucleus. Additionally, there
is some probability (dominant for light and medium-heavy nuclei) that the incident
hadron has no interactions prior to the point of DY pair production. In accordance
with the above considerations, the mean quark path in the nucleus can be written as \cite{Johnson:2001xf}:
\begin{eqnarray}
\label{eq:eloss_3}
& & \langle L\rangle = (1-W_0)\,\frac{\sigma^\text{hN}_\text{in}}{\text{A}}
\int \text{d}^2b\,\int\limits_{-\infty}^{\infty} \text{d}z_2\,
\rho_\text{A}(b,z_2) \nonumber \\
& &\times \int\limits_{-\infty}^{z_2} \text{d}z_1\,
\rho_\text{A}(b,z_1)\,(z_2-z_1) \, \nonumber \\
& &\times \text{exp}\left[-\sigma^\text{hN}_\text{in}
\int\limits_{-\infty}^{z_1} \text{d}z\,\rho_\text{A}(b,z)\right]\ .
\end{eqnarray}
The exponential factor requires that there is no inelastic interaction of the 
beam hadron prior to the point $z_1$. $W_0$ is the probability of no inelastic interaction 
of the beam hadron in the nucleus prior to the DY reaction, which can be written as
\begin{equation}
W_0=\frac{1}{\text{A}\,\sigma_\text{in}^\text{hN}}\int \text{d}^2b\,\left[
1-\e^{-\sigma_\text{in}^\text{hN}\,T(b)}\right] =
\frac{\sigma_\text{in}^\text{hA}}{\text{A}\,\sigma_\text{in}^\text{hN}}\ .
\end{equation}

The corresponding probability distribution in $L$ is given by expression 
\cite{Johnson:2001xf}:
\begin{equation}
\label{eq:wL}
W(L) = W_0\,\delta(L)\ +\ W_1(L)\ ,
\end{equation}
where
\begin{eqnarray}
  &&W_1(L) =
  \frac{\sigma^\text{hN}_\text{in}}{\text{A}}
  \int \text{d}^2b\,\int\limits_{-\infty}^{\infty} \text{d}z_2\,
  \rho_\text{A}(b,z_2) 
  \int\limits_{-\infty}^{z_2} \text{d}z_1\,
  \rho_\text{A}(b,z_1)\,\nonumber \\
  && \times \delta(z_2-z_1-L) \,
  \text{exp}\left[-\sigma^\text{hN}_\text{in}
    \int\limits_{-\infty}^{z_1} \text{d}z\,\rho_\text{A}(b,z)\right]
\end{eqnarray}

The HARDPING procedure to simulate quark energy loss uses two 
parameters: energy loss rate $-\text{d}E/\text{d}z$ (free parameter), and length of
a quark path $L$ in a nucleus, which is calculated according to 
\begin{figure}[htb]
\centering
\includegraphics[width=.9\hsize]{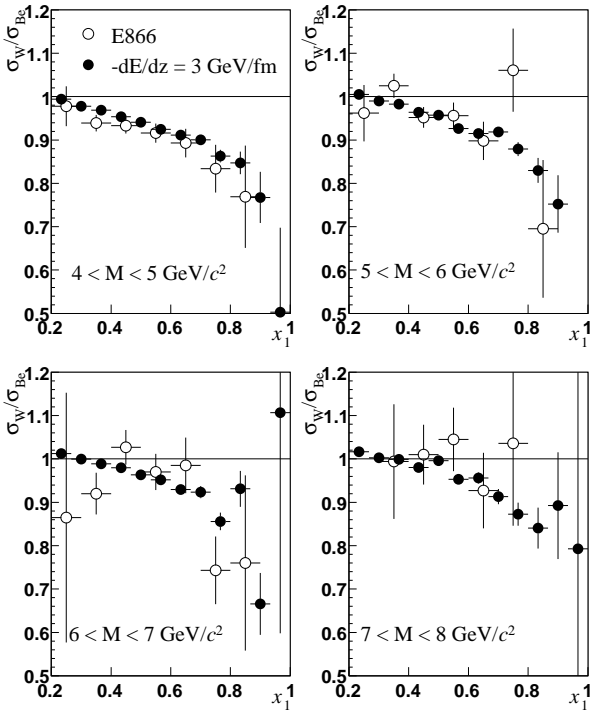}
\caption{Ratio of the cross sections for Drell-Yan events versus $x_1$ calculated for 
optimal value of the parton energy loss ratio $-\text{d}E/\text{d}z=3$~GeV/fm. $M$ is the invariant 
mass of produced DY pairs.}
\label{fig:longitudinal}
\end{figure}
the distribution eq.~(\ref{eq:wL}). 
The results of HARDPING calculations with energy loss rate 
$-\text{d}E/\text{d}z=3$~GeV/fm are presented in fig.~\ref{fig:longitudinal}, which is 
the DY cross section ratios for W to Be as functions of $x_1$ (the light-cone
momentum fraction of the incident proton carried by the produced DY pair) for various intervals
of the invariant mass $M$ of produced DY pairs. We calculated shadowing for the mean value of mass
calculated for each interval as $\sqrt{\langle M^2 \rangle}$.
One may expect a substantial scale dependence of
nuclear structure functions. However at considering ranges of $x_1 \ge 0.3$ and 
$4 \le M \le 8$~GeV/$c^2$ and at our statistical accuracy this dependence is not seen.
From comparison with the experimental data, it is found that our calculated results
are in good agreement with the Fermilab E866 data.

\section{Conclusion}
\label{sect:concl}
In this work the analysis of influence of the multiple soft rescatterings and the
energy loss of fast quarks effects on DY pairs production in proton-nucleus collisions were 
presented. In order to perform such an analysis we developed a Monte Carlo
event generator HARDPING, which extends HIJING program to the mentioned effects.

The analysis has shown that the proper treatment of multiple soft rescatterings of 
a quark of the incoming ha\-d\-ron allows to significantly improve the agreement 
between the simulated and experimental data comparing to the original HIJING. The shape
of the ratio $\sigma_\text{W}/\sigma_\text{Be}$ is very sensitive to variations of the mean 
momentum of the parton inside the valon $\langle k_\text{q}^\text{hard}\rangle$
and the cross section of the inelastic quark-nucleon interactions $\sigma$.

The simulation of the $x_1$-dependence of the ratio $\sigma_\text{W}/\sigma_\text{Be}$ showed
that consideration of the quark energy loss effect improves agreement with the 
experimental data. Original HIJING does not take into account this effect, which
leads to contardiction with the data.

However one should be cautious applying the results of this analysis to higher energies of RHIC and LHC,
because at RHIC and LHC energies coherence effects are important. The reason is that coherence length 
$l_c$ becomes large at high energies comparing to the typical internucleon separation, hence the projectile 
interacts coherently with individual nucleons. Moreover, the initial-state energy loss effect is not 
really important at the energies of RHIC and LHC. Thus one can disregard energy loss and test models 
of shadowing by direct comparison to data.

\begin{acknowledgement}
We are grateful to Yuli Shabelsky and Andrey Ivanov for their inspiring and clarifying discussions, 
and to V.B. Gavrilov, P.I. Zarubin, M.B. Zhalov, L.I. Sarycheva,
G. Feofilov, A.V. Khanzadeev for their helpful comments.
\end{acknowledgement}

\bibliographystyle{h-physrev3}
\bibliography{rmm_paper}

\begin{thebibliography}{10}

\bibitem{Arsene:2003yk}
BRAHMS, I.~Arsene {\em et~al.},
\newblock Phys. Rev. Lett. {\bf 91}, 072305 (2003), nucl-ex/0307003.

\bibitem{Adler:2003au}
PHENIX, S.~S. Adler {\em et~al.},
\newblock Phys. Rev. {\bf C69}, 034910 (2004), nucl-ex/0308006.

\bibitem{Back:2003qr}
PHOBOS, B.~B. Back {\em et~al.},
\newblock Phys. Lett. {\bf B578}, 297 (2004), nucl-ex/0302015.

\bibitem{Adams:2003kv}
STAR, J.~Adams {\em et~al.},
\newblock Phys. Rev. Lett. {\bf 91}, 172302 (2003), nucl-ex/0305015.

\bibitem{Cronin:1974zm}
J.~W. Cronin {\em et~al.},
\newblock Phys. Rev. {\bf D11}, 3105 (1975).

\bibitem{Aggarwal:1998vh}
WA98, M.~M. Aggarwal {\em et~al.},
\newblock Phys. Rev. Lett. {\bf 81}, 4087 (1998), nucl-ex/9806004.

\bibitem{Drell:1970wh}
S.~D. Drell and T.-M. Yan,
\newblock Phys. Rev. Lett. {\bf 25}, 316 (1970).

\bibitem{Gyulassy:1994ew}
M.~Gyulassy and X.-N. Wang,
\newblock Comput. Phys. Commun. {\bf 83}, 307 (1994), nucl-th/9502021.

\bibitem{Sjostrand:1993yb}
T.~Sjostrand,
\newblock Comput. Phys. Commun. {\bf 82}, 74 (1994).

\bibitem{Efremov:1985cu}
A.~V. Efremov, V.~T. Kim, and G.~I. Lykasov,
\newblock Sov. J. Nucl. Phys. {\bf 44}, 151 (1986).

\bibitem{Johnson:2001xf}
M.~B. Johnson {\em et~al.},
\newblock Phys. Rev. {\bf C65}, 025203 (2002), hep-ph/0105195.

\bibitem{Vasilev:1999fa}
FNAL E866, M.~A. Vasiliev {\em et~al.},
\newblock Phys. Rev. Lett. {\bf 83}, 2304 (1999), hep-ex/9906010.

\bibitem{Aubert:1983xm}
European Muon, J.~J. Aubert {\em et~al.},
\newblock Phys. Lett. {\bf B123}, 275 (1983).

\bibitem{Arneodo:1996rv}
New Muon, M.~Arneodo {\em et~al.},
\newblock Nucl. Phys. {\bf B481}, 3 (1996).

\bibitem{Adams:1995is}
E665, M.~R. Adams {\em et~al.},
\newblock Z. Phys. {\bf C67}, 403 (1995), hep-ex/9505006.

\bibitem{Amaudruz:1995tq}
New Muon, P.~Amaudruz {\em et~al.},
\newblock Nucl. Phys. {\bf B441}, 3 (1995), hep-ph/9503291.

\bibitem{Geesaman:1995yd}
D.~F. Geesaman, K.~Saito, and A.~W. Thomas,
\newblock Ann. Rev. Nucl. Sci. {\bf 45}, 337 (1995).

\bibitem{Alde:1990im}
D.~M. Alde {\em et~al.},
\newblock Phys. Rev. Lett. {\bf 64}, 2479 (1990).

\bibitem{Gavin:1991qk}
S.~Gavin and J.~Milana,
\newblock Phys. Rev. Lett. {\bf 68}, 1834 (1992).

\bibitem{Eskola:1998df}
K.~J. Eskola, V.~J. Kolhinen, and C.~A. Salgado,
\newblock Eur. Phys. J. {\bf C9}, 61 (1999), hep-ph/9807297.

\bibitem{Hirai:2001np}
M.~Hirai, S.~Kumano, and M.~Miyama,
\newblock Phys. Rev. {\bf D64}, 034003 (2001), hep-ph/0103208.

\bibitem{Kopeliovich:2002yh}
B.~Z. Kopeliovich, J.~Nemchik, A.~Schafer, and A.~V. Tarasov,
\newblock Phys. Rev. Lett. {\bf 88}, 232303 (2002), hep-ph/0201010.

\bibitem{Levin:1981mv}
E.~M. Levin and M.~G. Ryskin,
\newblock Yad. Fiz. {\bf 33}, 1673 (1981).

\bibitem{Voloshin:1982ry}
S.~A. Voloshin and Y.~P. Nikitin,
\newblock JETP Lett. {\bf 36}, 201 (1982).

\bibitem{Anisovich:1985xd}
V.~V. Anisovich, M.~N. Kobrinsky, J.~Nyiri, and Y.~M. Shabelski,
\newblock Sov. Phys. Usp. {\bf 27}, 901 (1984).

\bibitem{Lykasov:1984uf}
G.~I. Lykasov and B.~K. Sherkhonov,
\newblock Yad. Fiz. {\bf 38}, 704 (1983).

\bibitem{Alaverdian:1976qh}
G.~B. Alaverdian, A.~V. Tarasov, and V.~V. Uzhinsky,
\newblock Sov. J. Nucl. Phys. {\bf 25}, 354 (1977).

\bibitem{Garvey:2002sn}
G.~T. Garvey and J.~C. Peng,
\newblock Phys. Rev. Lett. {\bf 90}, 092302 (2003), hep-ph/0208145.

\bibitem{Kopeliovich:1984bf}
B.~Z. Kopeliovich and F.~Niedermayer,
\newblock JINR-E2-84-834.

\end{thebibliography}

\end{document}